\documentclass[%
 reprint,
superscriptaddress,
 amsmath,amssymb,
 aps,
]{revtex4-2}

\usepackage{graphicx}
\usepackage{dcolumn}
\usepackage[colorlinks=false,hidelinks=true]{hyperref}
\usepackage[mathlines]{lineno}
\usepackage{xcolor}
\usepackage{subfigure}
\usepackage{siunitx}
\DeclareSIUnit\angstrom{\text {\AA}}
\usepackage{comment}

\usepackage{glossaries}
\newacronym{md}{MD}{molecular dynamics}
\newacronym{ml}{ML}{machine learning}
\newacronym{nn}{NN}{neuronal network}
\newacronym{mlip}{MLIP}{machine-learned interaction potential}
\newacronym{eam}{EAM}{embedded-atom model}
\newacronym{nep}{NEP}{neuro-evolution potential}
\newacronym{msd}{MSD}{mean-squared displacement}
\newacronym{rdf}{PCF}{pair-correlation function}
\newacronym{se}{SE}{Stokes-Einstein}

\definecolor{brickred}{rgb}{0.8, 0.25, 0.33}
\usepackage[colorinlistoftodos,textcolor=brickred,backgroundcolor=white,bordercolor=brickred]{todonotes}

\begin{document}

\title{Molecular Dynamics simulations of Al-Ti metallic alloy melts using a transferable machine-learning potential}

\author{Yuna Kato}
\affiliation{School of Materials and Chemical Technology, Institute of Science Tokyo, Tokyo, Japan}
\affiliation{Institute for Frontier Materials on Earth and in Space, German Aerospace Center, Cologne, Germany}
\author{J\"urgen Brillo}
\author{Dirk Holland-Moritz}
\author{Fan Yang}
\affiliation{Institute for Frontier Materials on Earth and in Space, German Aerospace Center, Cologne, Germany}
\author{Thomas C. Hansen}
\affiliation{Institut Laue-Langevin (ILL), 38042 Grenoble, France}
\author{Thomas Voigtmann}
\author{Linnea Heitmeier}
\affiliation{Institute for Frontier Materials on Earth and in Space, German Aerospace Center, Cologne, Germany}
\affiliation{Institute for Theoretical Physics, Heinrich-Heine-Universit\"at D\"usseldorf, 40225 D\"usseldorf, Germany}
\email{linnea.heitmeier@dlr.de}

\date{\today}

\begin{abstract}
We investigate the structural and dynamical properties of binary aluminum--titanium liquid metallic alloys, as a function of temperature and composition. We make use of \gls{md}-simulations, using a transferable machine-learning potential developed by Song \textit{et~al.} [Nature Communications \textbf{15}, 10208 (2024)], and compare our results to experimental data.
Although this potential was initially trained on solid properties, we find good agreement between the experimental data and the simulation results for the liquid state.
The excess volume and compositional changes of the structure are captured
well by the machine-learned potential. The simulation allows to
disentangle local packing from chemical-ordering effects; the latter are
found to be weak in Al-Ti.
Dynamical quantities like the viscosity and the diffusion coefficients are
also discussed.
\end{abstract}

\glsresetall

\maketitle

\section{Introduction}
As alloys are produced from their melts, exact knowledge of the thermophysical
properties of the melts is important
\cite{brillo2016thermophysical, lv2011thermophysical, terekhov2023application}. 
Owing to their low density and high tensile strength, Al-based alloys play a key role in  aeronautic and automotive applications \cite{li2023development, alghannam2023investigation, davis1993aluminum, bhat2018aerospace}. 
The binary aluminum alloys where Al is combined with a semi-noble
metal provide good reference systems, and thus have been the focus of experimental and computational studies: for example Al-Ni and Al-Fe \cite{plevachuk2007density},
Al-Ag and Al-Cu \cite{brillo2008density}, Al-Au \cite{peng2015structural, sheng2011highly}, or Al-Ti \cite{zope2003interatomic, li2024review}.
Especially Al-Ti is a technologically relevant material due to its high melting temperature, strong chemical bonding, and technological relevance in lightweight, high-strength applications.

\Gls{md} simulations access the relevant length- and time scales to predict
thermophysical and transport properties of melts,
in particular as input quantities for further coarse-grained
simulations of material properties.
Due to its high technological relevance, a large body of work exists
on \gls{md} simulations of solid Ti-Al-based alloys \cite{li2024review};
the molten state is less well covered.
However, all these simulations
rely on ultimately empirical interaction potentials between the
atoms to describe a specific system.
The search for a suitable interaction model is generally laborious and
time consuming. Traditional approaches include the \gls{eam} or similar
empirical fits to force fields determined in quantum-mechanical
calculations.
Specifically for the Ti-Al system, \Citeauthor{zope2003interatomic}
\cite{zope2003interatomic} developed an \gls{eam} potential that they gauged
against a number of solid-state properties of the solid phase.
\Citeauthor{horbach2009ti} \cite{horbach2009ti} used this potential
for Ti in the liquid state but found that the potential had to be adjusted
in order to accurately describe the density and diffusion in the liquid.

A more modern approach is to use \gls{ml} methods for the task of
approximating the interparticle forces as functions of the atomic positions.
Typically \emph{ab~initio} simulation data is used to train
a \gls{nn} representing the interaction potential, including also
solid-state experimental properties.
This approach increasingly becomes attractive in order to calculate
the properties of alloy melts
\cite{fellman2025fast, sandberg2024homogeneous, zhao2024general, rahman2025machine}.
In principle, such \glspl{mlip} can be transferrable: once trained
on a sufficiently large training set, a suitably large \gls{nn}
has the capability of describing not just a specific alloy but rather
a larger set of materials.
However, the performance in comparison to experimental data related
to the melt remains to be checked, since also a \gls{mlip}
can only be as good as the underlying ab initio calculation.

In the present study we investigate the predictive capabilities
of a recent proposal of a very broadly transferable \gls{mlip},
the \gls{nep} potential NEP89 \cite{song2024general,liang2025nep89}, for the dynamics of Al-Ti
alloy melts.
The model has been trained on a large basis of data, and models of the
same architecture have demonstrated
success in the simulation of nanostructures \cite{liu2025temperature},
diamonds \cite{zhang2025thermal},
and HCP-zirconium \cite{jia5225377temperature}.
By performing \gls{md} simulations using NEP89 for Al-Ti melts
and comparing key thermophysical properties and transport coefficients
with the available experimental data, we provide a stringent test for
the transferability of the potential in an application-relevant context.
We demonstrate that the \gls{nep}, although not specifically trained on
the liquid-state regime that we are interested in here, performs well
and better than previous empirical potentials.

We note that recently, \Citeauthor{zhai2023ml} \cite{zhai2023ml}
performed dedicated training of a \gls{mlip} for the Ti-Al binary alloy
both in the solid and the liquid state, focusing on concentrations
$x_\text{Al}\approx0.5\approx x_\text{Ti}$ and on static properties.
While in principle, one expects such \glspl{mlip} to be transferrable
to different compositions, this has not been discissed in Ref.~\cite{zhai2023ml}.
Our study thus fills a gap by addressing specifically composition
dependence and also dynamical transport coefficients in the melt.

The paper is structured as follows:  
In Sec.~\ref{sec:methods} we briefly summarize the simulations and
experimental methods relevant for this paper.
Our results are presented in Sec.~\ref{sec:results}: we start by
investigating the static thermophysical properties, comparing the
composition-dependent mass density, molar volume, and the molar excess volume
to experimental data
(Sec.~\ref{sec:results_densities}).
A discussion of the microscopic structure, in terms of the
static structure factors, the related \glspl{rdf},
and coordination numbers follows (Sec.~\ref{sec:structurefactor});
here, experimental data are available and are being compared to in the pure Ti melt.
We then turn to dynamical quantities, i.e., the mass-transport coefficients:
the viscosity (Sec.~\ref{sec:viscosity}) is compared to experimental
data available for Al-rich melts, while for self- and interdiffusion
coefficients (Sec.~\ref{sec:results_diffusioncoefficient}), reliable
experimental data is still lacking so that our \gls{md} simulation
results serve as predictions of the experimentally validated model.
We conclude in Sec.~\ref{sec:conclusion}.

\section{Methods}
\label{sec:methods}

\subsection{Computer simulations}
\label{sec:methods_simulations}
Molecular-dynamics simulations are performed using the open-source software \textsc{lammps} \cite{thompson2022lammps}. As interatomic potentials, we employ the models developed by \Citeauthor{song2024general} \cite{song2024general} and
\Citeauthor{liang2025nep89} \cite{liang2025nep89}.
Unless stated otherwise, all simulations are carried out with N=10000 atoms and periodic boundary conditions in all three spatial directions. The equations of motion are integrated using the standard velocity-Verlet algorithm with a time step of $1$ fs. 

Each simulation run is first equilibrated at 2000 K for $500$ ps and then cooled down to the desired temperature in the NPT
ensemble (constant particle number, zero external pressure,
and linearly decreasing temperature) for $10$ ps. 
After further equilibration at the desired temperature, production runs are performed for a time of 200 ps. 
Temperatures
considered in this work are $T=1550$, $1650$, $1720$, $1800$, $1873$, $1950$, $2000$, $2100$, and $2200\ \mathrm{K}$,  
Compositions are specified as Aluminium mole fractions $x_{Al}$  and include
$x_{Al}$= 0.0, 0.1, 0.2, 0.3, 0.4, 0.5, 0.6, 0.7, 0.8, 0.9, 1.0.

The density is calculated as  
\begin{equation}
   \rho =  \frac{N (x_{\text{Al}}m_{\text{Al}}+ x_{\text{Ti}}m_{\text{Ti}})}{V}
\end{equation}
where $x_\alpha$ ($\alpha\in\{\text{Al},\text{Ti}\}$) is the number
concentration (molar fraction) of the respective element, $m_\alpha$ is the
corresponding atomic
mass, $N$ is the number of particles in the box, and $V$ is the volume of the system.
For the atomic masses, we used values of $m_{\text{Ti}} = \qty{47.86}{u}$ and $m_{\text{Al}} = \qty{26.98}{u}$.

The viscosity $\eta$ is determined through the Green-Kubo-relation, using the time-dependent stress-autocorrelation function of the off-diagonal stress tensor elements,
\begin{equation}
    \eta = \frac{V}{k_B T}\int_0^\infty dt' \langle P_{j}(0) P_{j}(t')\rangle_{j\in\{(xy), (xz), (yz)\}} 
\end{equation}
where $T$ is the temperature and $k_B$ is the Boltzmann constant.
The self-diffusion coefficients $D_\alpha$ were obtained from the long-time behavior of the \gls{msd} according to the Einstein relation,
\begin{equation}
    D_\alpha = \frac{\langle |\mathbf{r}_\alpha(t) - \mathbf{r}_\alpha(0)|^2 \rangle}{6 t},
\end{equation}
where $\mathbf{r}_\alpha(t)$ is the position of a given particle of species $\alpha$ at time $t$ (and the average is performed over all particles of the species).

The inter-diffusion coefficient $D_{cc}$ is calculated following the procedure
described in Ref.~\cite{horbach2007self}, from a corresponding Einstein
relation,
\begin{multline}
\label{eq:interdiff_msd}
D_{cc}= \lim_{t\to\infty} \left( 1+\frac{m_\text{Al} x_\text{Al}}{m_\text{Ti} x_\text{Ti}}\right)^2 \times\\ \times N x_\text{Al} x_\text{Ti} \Phi \frac{\langle (\mathbf{R}^{(\text{Al})}(t)-\mathbf{R}^{(\text{Al})}(0))^2\rangle}{6t}\,,
\end{multline}
where $\mathbf{R}^{(\text{Al})}=(1/N_\text{Al})\sum_{i=1}^{N_\text{Al}}\mathbf{r}_{\text{Al},i}$
is the center-of-mass coordinate of all Al particles.

The thermodynamic prefactor
$\Phi={x_\text{Al} x_\text{Ti}}{S^{-1}_{cc}(q\to 0)}$ is determined by
extrapolating $S_{cc}$ to $q \to 0$ in the simulation data.
The appearance of such a thermodynamic prefactor is the signature of
a collective transport process, and it allows to separate the thermodynamic
mixing effects from purely kinetic ones. The latter are expressed through
the Onsager coefficient $L$, defined through the relation
\begin{equation} D_{cc}=L\Phi\,. \end{equation}

The relation between the self-diffusion and interdiffusion processes is a
topic of long-standing interest \cite{horbach2007self}. In particular,
the empirical Darken equation \cite{darken1948diffusion} suggests the
relation
\begin{equation}
\label{eq_interdiff_darken}
  D_{cc} \approx L_\text{Darken}\Phi =
    \Phi (x_\text{Al} D_\text{Ti} +x_\text{Ti} D_\text{Al} )\,,
\end{equation}
and deviations from the Darken approximation indicate genuine collective processes
that contribute to interdiffusion.

The \gls{rdf} was calculated by
\begin{equation}\label{eq:grab}
g_{\alpha\beta}(r) = \frac{1}{4\pi r^2 \Delta r n N} \sum_{i=1}^{N_\alpha}\sum_{j=1,j\ne i}^{N_\beta}  \chi_{[r, r+\Delta r)}(r_{ij})
\end{equation}
where $n=N/V$ is the number density, the sums run over particles of the
$\alpha$ and $\beta$ species respectively,
$\Delta r$ is the width of the bin, and $\chi(x)$ is 1 when $x \in [r, r+\Delta r)$ and 0 otherwise.  
The function quantifies the probability of finding a particle of species $\beta$ at a distance $r$ from a reference particle of species $\alpha$, normalized by the average density. In our simulations, we made use of the LAMMPS in-build method to calculate $g_{\alpha\beta}(r)$ \cite{lammps_rdf}. The total \gls{rdf}
$g(r)$ is obtained when no distinction of the species is made
in Eq.~\eqref{eq:grab}.

The coordination number (CN) was evaluated using two different approaches following the calculation of the \gls{rdf}.
The first method is based on the integration of the first peak of $g(r)$,
\begin{equation}
CN = 4 \pi \rho \int_{0}^{r_c} r^2 g(r) \, dr,
\label{eq:CN_1}
\end{equation}
where $\rho$ is the atomic number density and $r_c$ is the cutoff distance corresponding to the first minimum of $g(r)$. 
This integration yields the average number of neighboring atoms within the first coordination shell.
The second method utilizes atomic configurations obtained from the simulations. 
For each atom $i$, the number of neighboring atoms $j$ within the cutoff radius $r_c$ was directly counted as
\begin{equation}
CN_i = \sum_{j \neq i} \Theta(r_c - r_{ij}),
\label{eq:CN_2}
\end{equation}
where $r_{ij}$ is the distance between atoms $i$ and $j$, and $\Theta$ is the Heaviside step function. 
The average coordination number was then obtained by averaging over all atoms and sampled configurations.

We also calculate the static structure factor, assuming that the system is isotropic:
\begin{equation}\label{eq:sabdef}
    S_{\alpha\beta}(q)= \frac{1}{N} \biggl \langle \sum_{i=1}^{N_\alpha} \sum_{j=1}^{N_\beta} e^{-i q |r_{\alpha,i}-r_{\beta,j}|} \biggr \rangle\,.
\end{equation}
In our calculation, for each $q$, we choose $120$ different directions and sum up the atomic distances directly. For efficiency reasons, we only sum up those distances, which are smaller than a cutoff-distance. The method of calculating
$S_{\alpha\beta}(q)$ directly from the particle positions suffers from
somewhat higher statistical noise, but is much less prone to spurious
osciallations induced by a finite-range Fourier transform of the
$g_{\alpha\beta}(r)$.
Note the convention
\begin{equation}\label{eq:hq}
  S_{\alpha\beta}(q)=\delta_{\alpha\beta}x_\alpha
  +n x_\alpha x_\beta h_{\alpha\beta}(q)
\end{equation}
with the number density $n$,
and the total correlation function $h_{\alpha\beta}(q)$ that is essentially
the Fourier transform of the \gls{rdf},
$h_{\alpha\beta}(r)=g_{\alpha\beta}(r)-1$.
This implies $S_{\alpha\beta}(q\to\infty)=\delta_{\alpha\beta}x_\alpha$.
The often used Ashcroft-Langreth structure factors are obtained
from $S_{\alpha\beta}(q)/\sqrt{x_\alpha x_\beta}$.

\subsection{Experiments}
\label{sec:methods_experiments}
As liquid Ti alloys are chemically highly reactive, containerless processing techniques, such as electrostatic or electromagnetic levitation are the methods of choice in order to avoid pollution of the samples due to reactions with a substrate or a crucible. The data of density, molar volume and the structure factor are measured containerlessly. 

Density and molar volume were obtained in Ref.~\cite{BrilloIJT2025} using electromagnetic levitation \cite{MohrReviewIJT2024} and a shadowgraph technique sometimes referred to as optical dilatometry \cite{BrilloIJT2024}. In this technique, a shadow of the levitating droplet is projected from the side onto the sensor of a camera. After recording a series of images, the sample volume is determined by integration of the time-averaged edge curve. Rotational symmetry regarding the vertical axis is assumed for the equilibrium shape.
The uncertainty of this method is $\pm1.5\%$

The experimental structure factor $S(q)$ was measured by neutron scattering combined with electromagnetic levitation \cite{BrilloIJT2024,A,B}.
The neutron diffraction experiments were carried out at the two-axis diffractometer D20
of the Institut Laue-Langevin (ILL) in Grenoble, France, using a monochromatic neutron beam with a wavelength of $\qty{0.81}{\angstrom}$ (pure Ti)
and $\qty{0.94}{\angstrom}$ (TiAl).
The diffractometer is equipped with a large area linear curved position-sensitive detector that covers 
an angular range of $\qty{153.6}{\degree}$.
Systems of slits before and behind the sample make sure that 
the irradiated sample volume is as large as possible while, at the same time scattering contributions from the Al-window of the experimental 
chamber or from the levitation coil are avoided.
The static structure factors of the melts, $S(q)$, as function of the
momentum transfer, $q$, have been determined from the raw diffractograms
by applying the same data treatment as described in Ref.~\cite{A},
which includes background subtraction, the correction of self-absorption,
multiple scattering and of the effect of inelastic scattering
and normalization using a solid vanadium reference. Using this setup
structure factors of pure liquid Ti have been measured that have been published in Ref.~\cite{holland2007short}. In this work we present new results
on neutron diffraction experiments on melts of TiAl.
Within the formalism of Bhatia and Thornton \cite{C} the
measured total static structure factor $S(q)$ of a binary AB alloy melt
is composed of three partial static structure factors:
$S_{nn}(q)$, $S_{cc}(q)$, and $S_{nc}(q)$.
$S_{nn}(q)$ describes the topological short-range order of the liquid,
$S_{cc}(q)$ the chemical short-range order, and $S_{nc}(q)$
he correlation of number density and chemical composition. The relation between the total structure factor measured in a scattering experiment
and these partial structure factors is given by
\begin{equation}\label{eq:bt}
S(q)=\frac{{\bar b}^2}{\bar{b^2}}S_{nn}(q)
+\frac{(b_A-b_B)^2}{\bar{b^2}}S_{cc}(q)
+\frac{2(b_A-b_B)\bar b}{\bar{b^2}}S_{nc}(q)\,.
\end{equation}
Here $b_\alpha$ denote the coherent neutron scattering lengths of the atoms
of type $\alpha$, $\bar b=x_Ab_A+x_Bb_B$, and
$\bar{b^2}=x_Ab_A^2+x_Bb_B^2$.

Due to the negative coherent neutron scattering length of Ti
($b_\text{Ti}=-\qty{3.37}{\femto\meter}$) and the positive scattering
length of Al ($b_\text{Al}=\qty{3.45}{\femto\meter}$) the Ti$_{50.58}$
Al$_{49.42}$ alloy that is close to the equimolar composition corresponds to a zero-scattering alloy,
which means that the average scattering length $\bar b$ vanishes.
For this case the coefficients of $S_{nn}(q)$ and $S_{nc}(q)$
in Eq.~\eqref{eq:bt} vanish, while the coefficient of $S_{cc}(q)$
is $1/x_\text{Al}x_\text{Ti}$. Therefore a neutron scattering experiment
on the zero-scattering alloy Ti$_{50.58}$Al$_{49.42}$ directly gives $S_{cc}/x_\text{Al}x_\text{Ti}$.

The viscosity data shown in this work was measured by Takeda \cite{TakedaSatoITJ2025}
using the oscillating cup technique \cite{BrilloIJT2024}. Very good care has been taken here in order to avoid crucible reactions in the latter case.  
In the oscillating cup technique, the melt is inside a cylindrical crucible which is suspended by a torsion wire and performs angular oscillations which decay with time due to the inner friction of the liquid. The decay time is measured by an optical arrangement and the viscosity is calculated from it by numerically solving the Navier-Stokes equations for that problem. The uncertainty of this method can be up to $\pm20\%$.

\section{Results}
\label{sec:results}
\subsection{Densities}
\label{sec:results_densities}

\begin{figure}
    \centering
    \includegraphics[width=\linewidth]{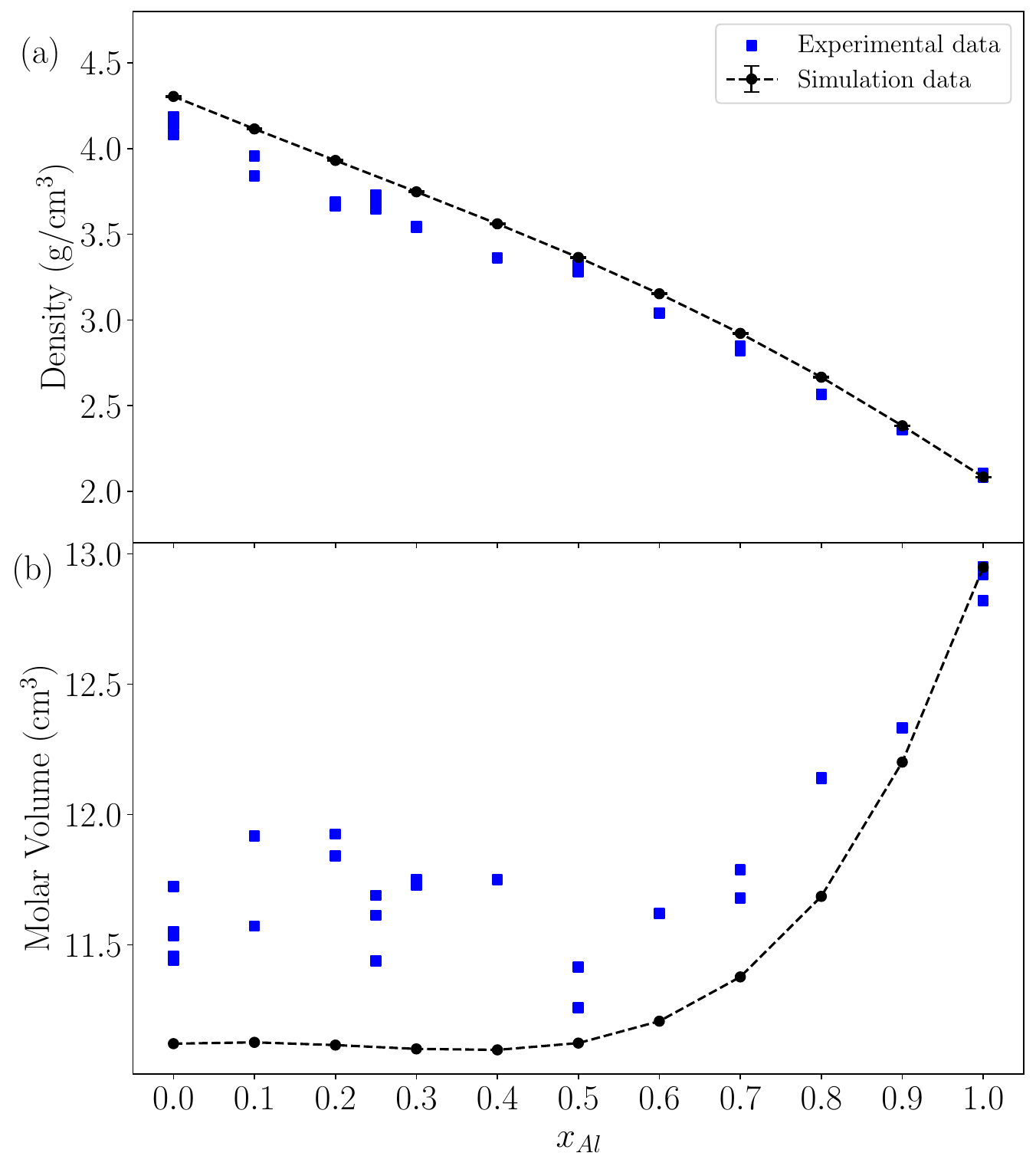}
    \caption{Panel a) shows the mass density of the Al-Ti melts at $T=\qty{1873}{\kelvin}$ as a function of Aluminium concentration. Black circle symbols joined by dotted lines are simulation data, blue squares represent experimental data from Ref.~\protect\cite{brillo2025thermophysical}.
    Panel b) shows the corresponding molar volume.
    \label{fig:vd_00} 
    }
\end{figure}

Figure~\ref{fig:vd_00} shows the mass density and the corresponding molar volume as a function of the composition, at fixed temperature of 1873 K. In general, the simulation results are in good agreement with experimental data. For the equimolar composition, the simulation data from Ref.~\cite{zhai2023ml} show a similar level of agreement. This highlights the high transferability of the \gls{nep} potential that is able to compete with the dedicated \gls{nn} trained on Al-Ti. A systematic overestimation of the density and corresponding underestimation of the molar volume is evident on the Ti-rich side. The discrepancy is on the order of $5\%$ and becomes noticeable for $x_\text{Al}\lesssim0.5$. Still the simulation captures well the fact that in this regime, the molar volume remains almost independent of concentration, while for $x_\text{Al}\gtrsim0.5$ the molar volume increases with increasing Al content noticeably.
We cannot fully rule out systematic errors in experiment, but the data is corroborated by multiple sources \cite{BrilloIJT2025,brillo2019normal}.
Note that also \gls{eam} potentials fail to accurately predict the density of Ti in its liquid state and have to be adjusted by empirical scaling factors \cite{horbach2009ti}. The \gls{nep} potential performs much better than this traditional epmirical potential.

\begin{figure}
    \centering
    \includegraphics[width=\linewidth]{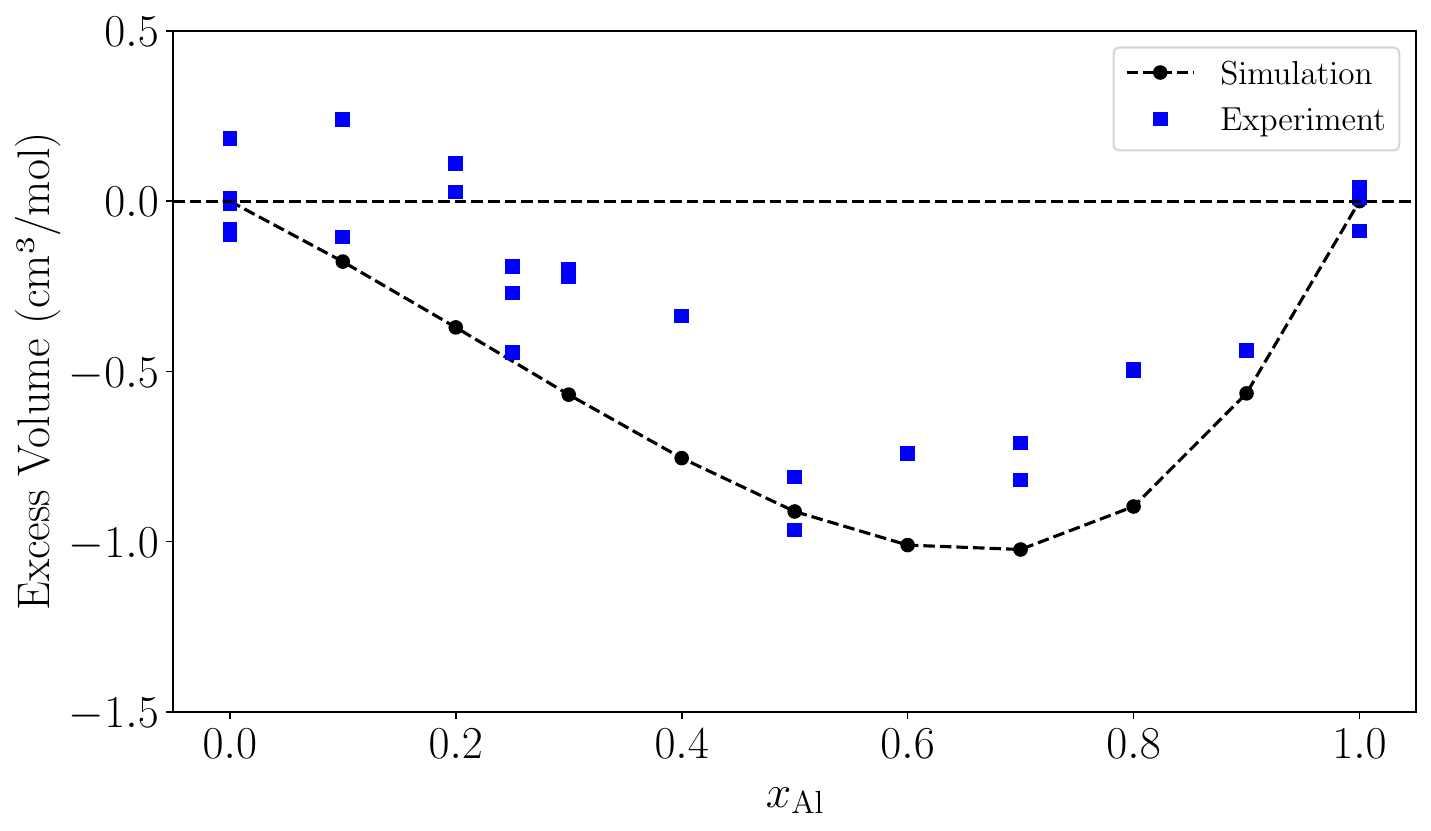}

    \caption{Excess volume as a function of Al concentration at fixed temperature $T=\SI{1873}{\kelvin}$. Circle symbols joined by dashed lines represent the simulation data, squares the experimental data from Ref.~\cite{brillo2025thermophysical}}.
    \label{fig:vd_11} 
\end{figure}

A convenient measure to quantify non-ideal mixing of the two components of the melt is the excess volume, defined as
\begin{equation}
V_{\text{excess}} = V - \left( x_{\text{Al}} V_{\text{Al}} + x_{\text{Ti}} V_{\text{Ti}} \right)\,,
\end{equation}
where $V$ is the molar volume of the mixture, and $V_\text{Al}$ and $V_\text{Ti}$, respectively, are the molar volumes of the pure systems. In essence, the excess volume measures the deviation from the ideal-mixing solution. 

Figure~\ref{fig:vd_11} compares the simulation results with available experimental data (obtained from the densities measured in containerless electromagnetic levitation using optical dilatometry \cite{BrilloIJT2024}).
While the data generally indicates negative excess volumes for
this binary system, some experimental data points at low Al concentrations have positive values;
from the scatter of the data in this concentration range
we estimate the experimental error to be on the order of $0.3 \text{ cm}^3/\text{mol}$.
Within the estimated experimental uncertainty, the simulation reproduces the general trend observed in experiments. Here, a negative excess volume is found for all compositions. A minimum in the excess volume occurs at a molar fraction of $x_\text{Al}\approx0.6\pm0.1$, in good agreement with experiment. 
The negative excess volume is also found in other Al-based alloy melts
\cite{peng2015structural}.

\subsection{Static structure}
\label{sec:structurefactor}

\begin{figure}
    \centering
    \includegraphics[width=\linewidth]{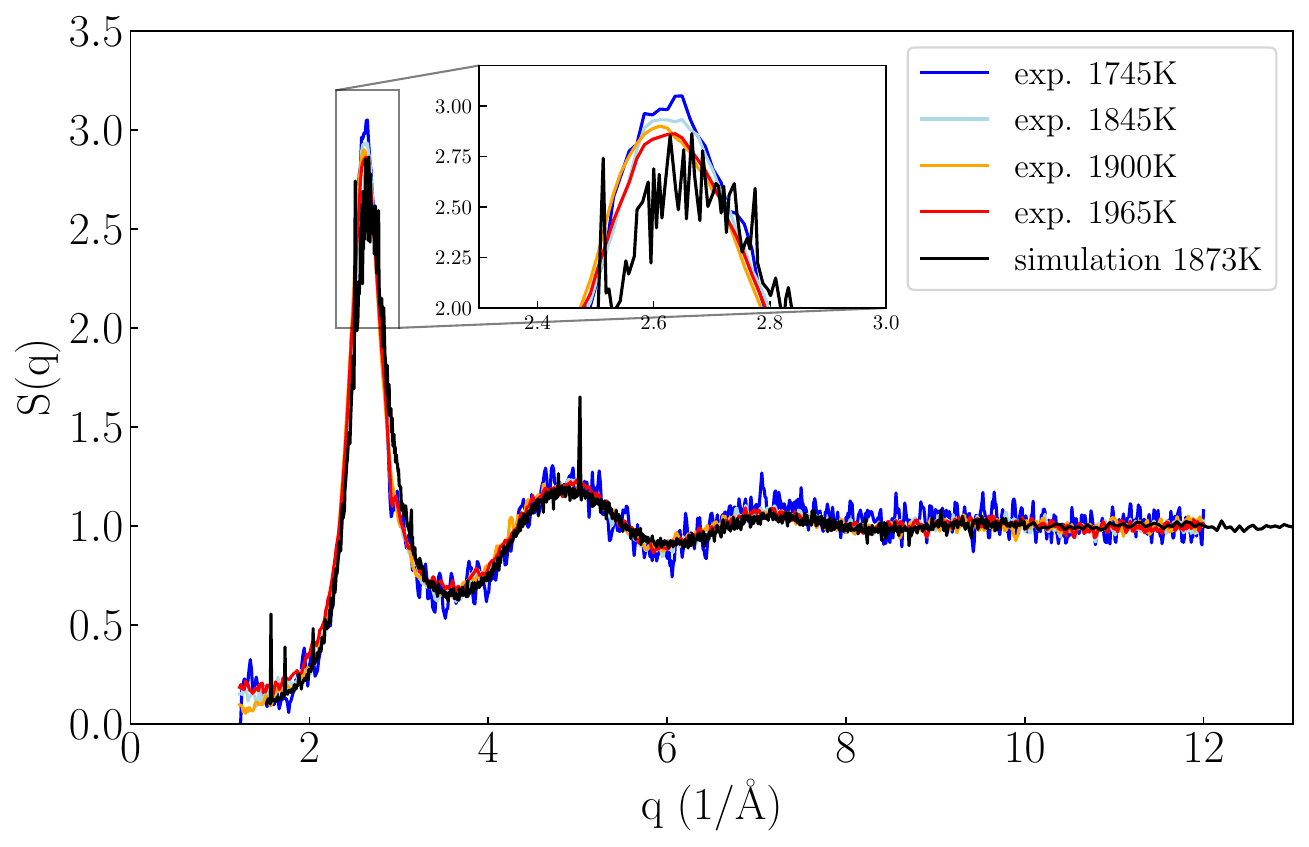}
    \caption{Static structure factor for pure Titanium. Experimental data taken from \cite{holland2007short}}
    \label{fig:sq}
\end{figure}

The microscopic structure of the liquid is characterized by the
partial pair-correlation functions $g_{\alpha\beta}(r)$ or, equivalently,
the partial static structure factors. While essential to understand the
chemical ordering and entropic mixing effects, the partial structure
functions are not always readily accessible in experiment; often,
only linear combinations (given by the atomic scattering lengths) are
obtained through X-ray or neutron scattering.

For the pure Ti system, neutron-diffraction
experiments have determined $S(q)$ \cite{holland2007short}.
Figure~\ref{fig:sq} shows a comparison at different temperatures to our
simulation results.
Across the temperature range accessed in the experiment, $S(q)$ varies
only slightly; a variation that cannot be resolved in the plot.
We find quite accurate agreement between simulation and experiment,
with perhaps the exception of the large-$q$ oscillations that are
slightly more pronounced in the experimental data.
However, the most relevant regions for the liquid region, around the
first and second peaks in $S(q)$ is well resolved.

\begin{figure}
    \centering
    \includegraphics[width=\linewidth]{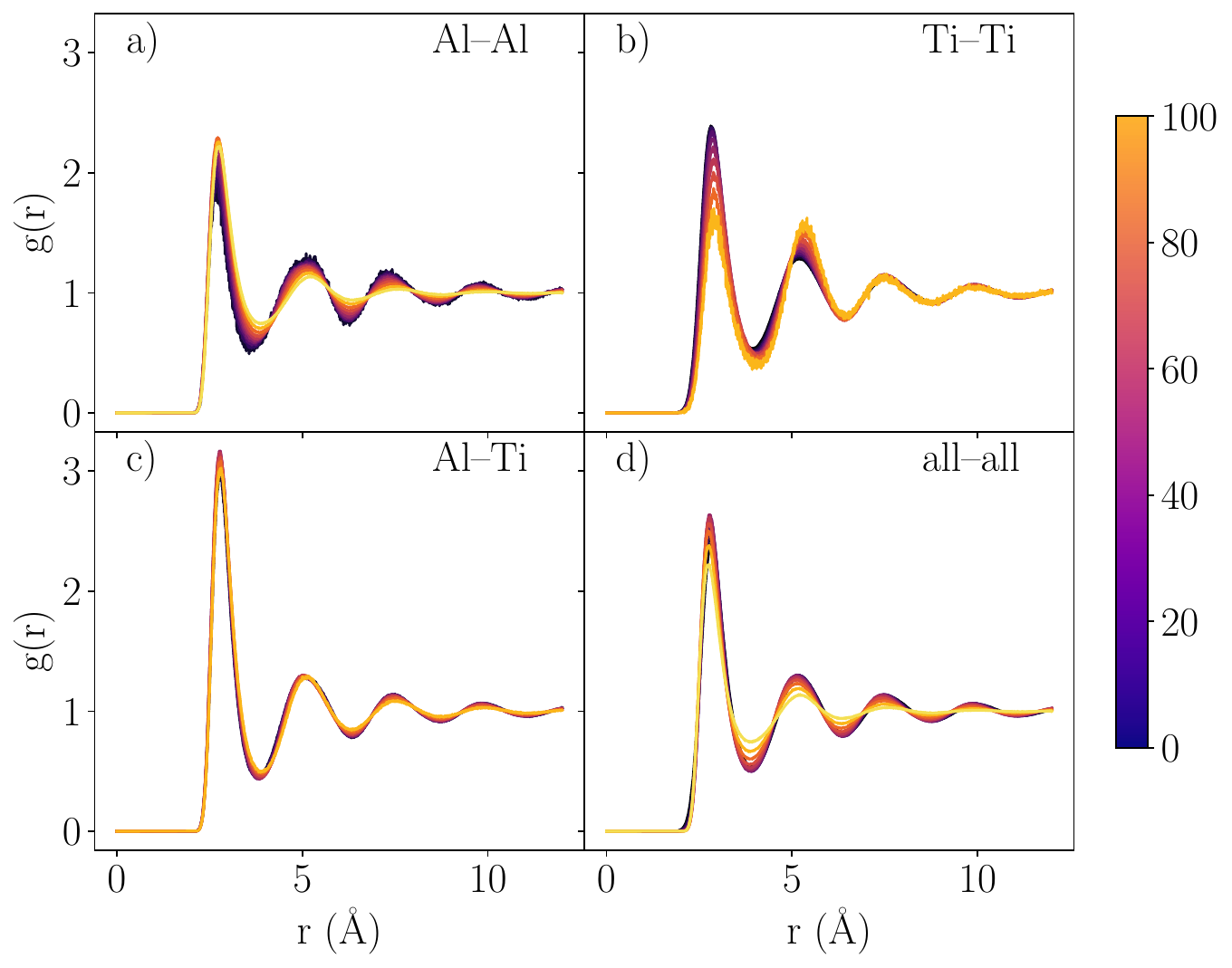}
     \caption{\label{fig:rdf}Partial and total pair-correlation functions $g(r)$ at fixed temperature $T=1873 K$. Color codes the Al concentration from dark ($x_\text{Al}=0$) to bright ($x_\text{Al}=1$).}
\end{figure}

Turning to the concentration-dependent microstructure, we show
the partial \glspl{rdf} in Fig.~\ref{fig:rdf};
here we focus on the composition change at fixed temperature of 1873 K.
For reference, the total $g(r)$ is also shown.
For this Al-Ti system, the atomic distances between Al and between Ti
atoms are quite similar, as signified by the very similar main peak positions
in the $g_{\alpha\beta}(r)$. As a result, the total structure function
changes little with composition; its peaks are slightly less pronounced
for the Al-rich systems than for the Ti-rich ones. This indicates that
at the temperature considered here, Ti is more strongly ordered by a small
amount; the expectation (that will be confirmed below) is thus that along this
isotherm, the Ti dynamics will be slower than the Al one.

The cross term $g_{\text{Al}\text{Ti}}(r)$ also shows little concentration
dependence and bears the same peak structure as the diagonal terms.
This indicates that chemical ordering in the Al-Ti liquid is not very
pronounced, and mixing occurs mainly by substitutional disorder (i.e.,
adding Al atoms to the Ti-rich melt mostly replaces Ti atoms in the
packing by Al ones, without altering the structure significantly).
This conclusion is in line with the results of \textit{ab initio}
\gls{md} simulations of Ti-Al melts \cite{feng2023aimd}.

\begin{figure}
    \centering
    \includegraphics[width=\linewidth]{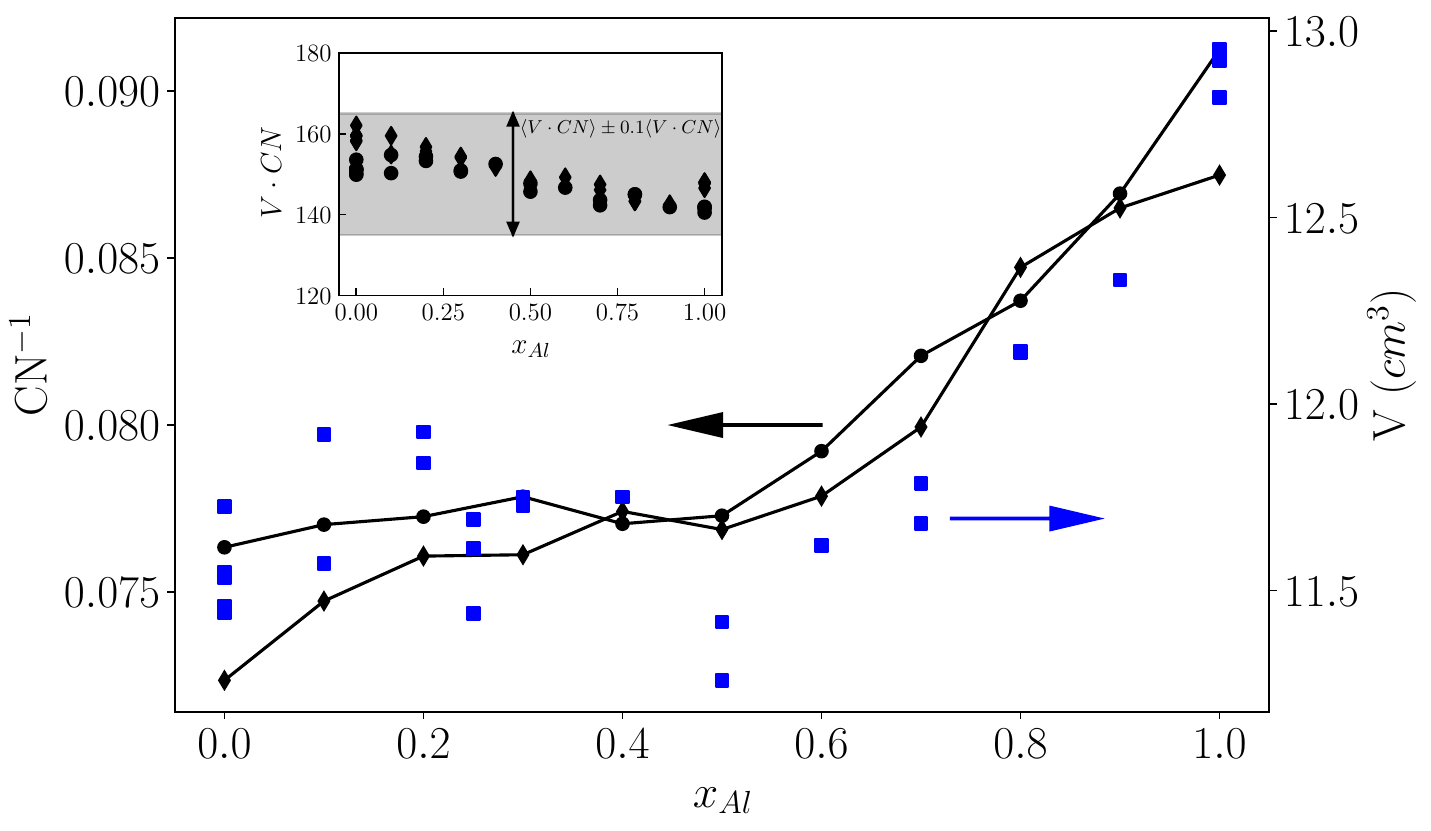}
    \caption{Inverse coordination number from simulations (black diamonds uses equation \eqref{eq:CN_1} and black circles uses \eqref{eq:CN_2}; left ordinate), and experimental molar volume (blue; right ordinate) the Al-Ti melts at $T=\qty{1873}{\kelvin}$ as a function of composition. Inset: product $CN\,V$ to demonstrate the similar composition dependence.}
    \label{fig:coordinationnumber}
\end{figure}

The $g(r)$ allow to define a coordination number of the local structure,
see Sec.~\ref{sec:methods_simulations}.
Since both Al and Ti are atoms of similar size, the inverse coordination numbers
should closely reflect the local molar volume, and indeed there is a
strong correlation between the two quantities,
as seen in Fig.~\ref{fig:coordinationnumber}.
Both Al and Ti show very similar coordination numbers, confirming the
observation of mostly substitutional mixing that was
discussed in the context of the $g_{\alpha\beta}(r)$.

\begin{figure}
    \centering
    \includegraphics[width=\linewidth]{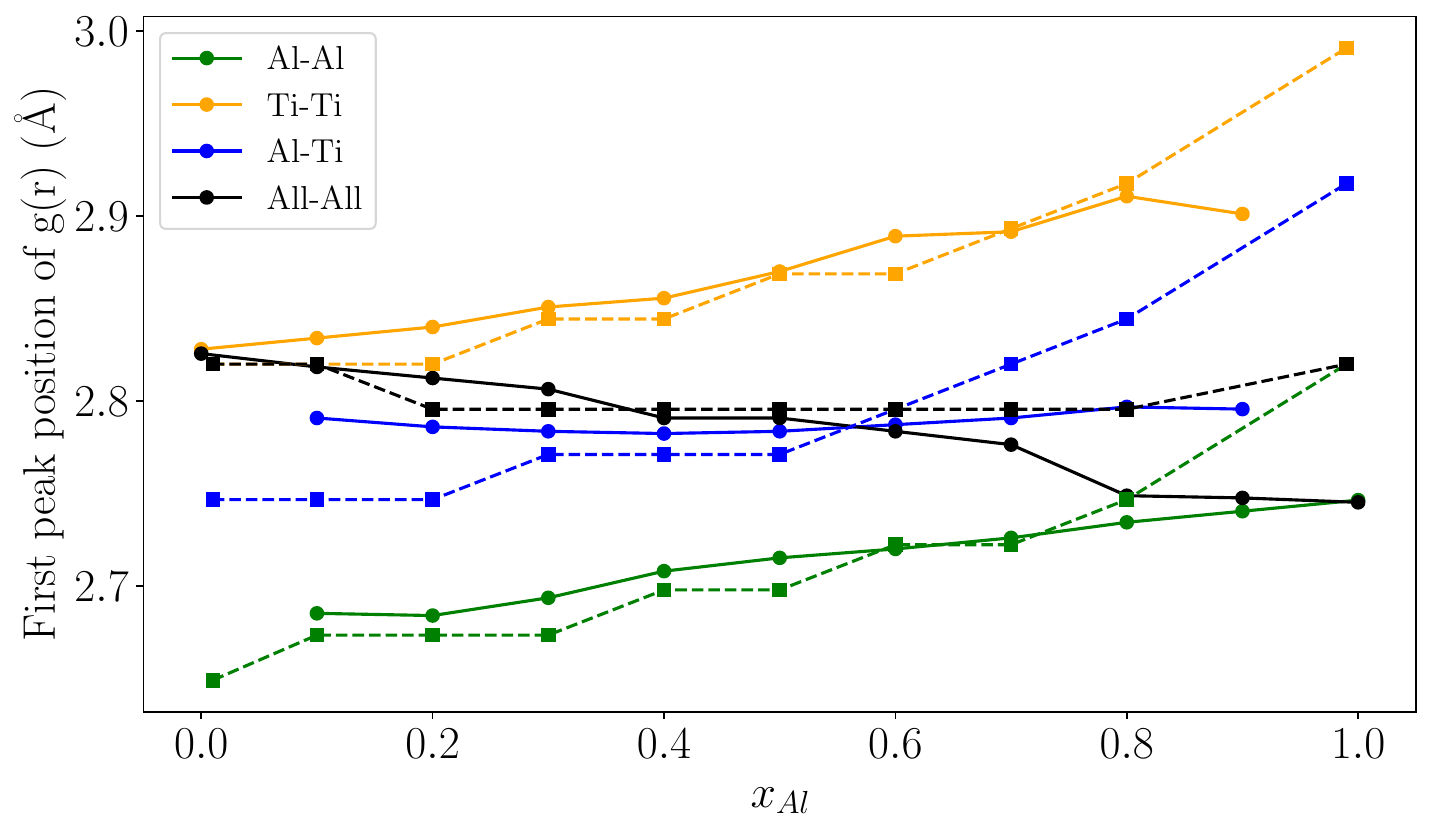}
    \caption{First-peak positions of pair the correlation functions against concentration ($T=\qty{1873}{\kelvin}$).
    Solid line: simulation results. Dashed lines: First-peak positions from a soft-sphere model, obtained within the HNC approximation.}
    \label{fig:rdf_01}
\end{figure}

A common model for mixing effects in metallic melts is that of hard spheres
of different sizes, where each atomic species is assigned an effective size
$\sigma_\alpha$.
One way of estimating this size is via the position of the first peak
in the pair distribution functions.
We show the composition dependence of these quantities in Fig.~\ref{fig:rdf_01}.
Note that for the hard-sphere model, the peak positions would be solely
given by the sizes of the spheres, and be independent of the mixture composition.
Instead, we observe a monotonous increase of both the AlAl and the TiTi
peak positions with increasing Al mole fraction. A similar trend has been seen in AlAu
and was interpreted as a local compression \cite{peng2015structural}.
For the Al-Ti melt studied here, the compositional changes in microstructure
are in fact well explained by a simple soft sphere model:
replacing the idealized hard-sphere interaction by a potential of the
form $V_{\alpha\beta}(r)=\epsilon(r/\sigma_{\alpha\beta})^{-n}$ with
$\sigma_{\alpha\beta}=(\sigma_\alpha+\sigma_\beta)/2$ and $n=12$,
the simulation model of Al-Ti can be well described (dashed lines
in Fig.~\ref{fig:rdf_01}).
These data were obtained from a numerical solution of the
Ornstein-Zernike equation with the hypernetted-chain (HNC) closure
\cite{verlet1968computer,hansen2013theory};
the soft-sphere model parameters were $\sigma_\text{Al}=\qty{2.7}{\angstrom}$,
$\sigma_\text{Ti}=\qty{2.896}{\angstrom}$ and $k_BT/\epsilon=0.4$.
We discuss the choice of particle diameters further on in connection
with the hydrodynamic radii extracted from the self-diffusion coefficients.
This choice provides a good match for low Al concentrations, at the cost
of strong deviations around pure Ti.
Also the soft-sphere model fails to accurately predict the entire $g(r)$.
Since hard and soft spheres can be seen as model systems with purely
entropic mixing, and the trend predicted by the soft sphere model for the
peak positions matches the simulation data
reasonably well for low Al concentrations. From this one might conclude that
chemical short-range order is not dominant to predict the first-shell
interatomic distances in the Ti-Al melts with
$x_\text{Al}\lesssim0.5$; but above that concentration, non-entropic
mixing effects can be expected to become more relevant.
Note that the hard-sphere model
is too simplistic to draw this conclusion, although it has often been
used as a theoretical reference \cite{peng2015structural}.

\begin{figure}
    \centering
    \includegraphics[width=\linewidth]{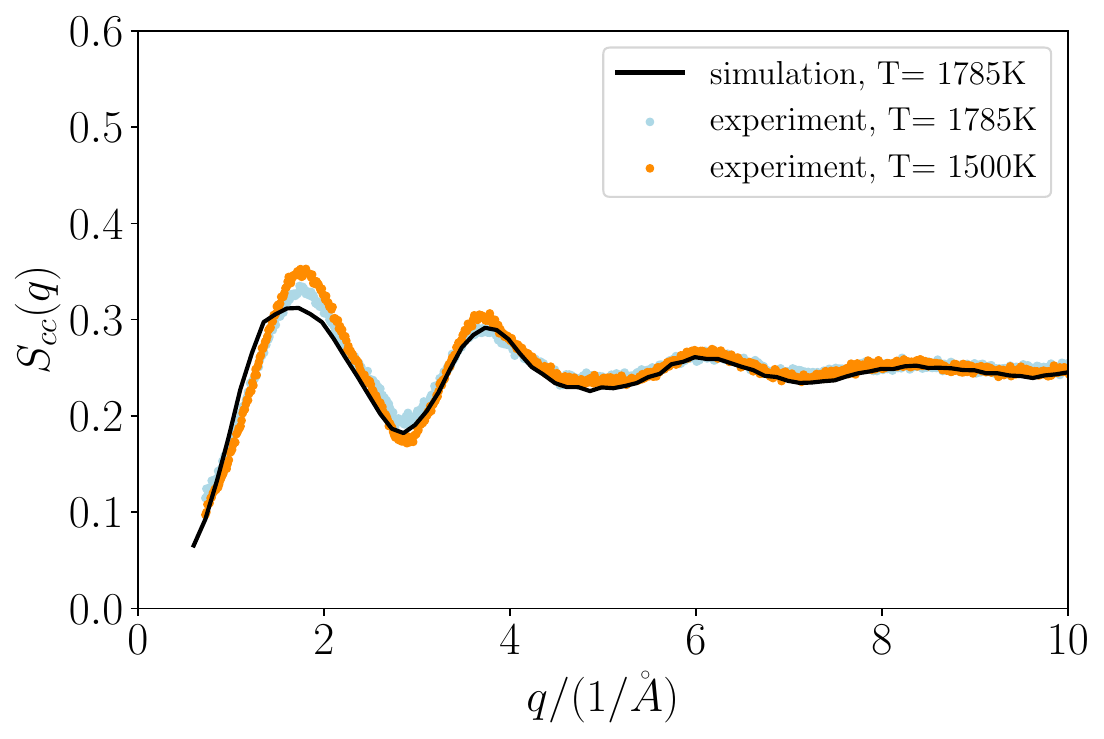}
    \caption{Concentration-concentration static structure factor for the
    equimolar Al-Ti melt at $T=\qty{1785}{\kelvin}$ (line).
    Symbols represent experimental data from a zero scatterer (Al$_{49.42}$Ti$_{50.58}$),
    at two different temperatures as indicated.}
    \label{fig:SCC}
\end{figure}

A more relevant view on chemical-ordering and mixing effects is provided by the
concentration-concentration static structure factor $S_{cc}(q)$.
In particular, its
$q\to0$ value also determines the thermodynamic factor for interdiffusion.
It is easily obtained from the partial structure factors in the simulation,
\begin{equation}
    S_{cc}(q) = x_{\text{Ti}}^2 S_{\text{AlAl}}(q) + x_{\text{Al}}^2
      S_{\text{TiTi}}(q) - 2 x_{\text{Al}} x_{\text{Ti}} S_{\text{AlTi}}(q)\,,
\end{equation}
where $S_{cc}(q\to\infty)=x_\text{Al}x_\text{Ti}$ follows from the convention
we use for $S_{\alpha\beta}(q)$, cf.\ Eq.~\eqref{eq:sabdef}.
We show in Fig.~\ref{fig:SCC} an exemplary simulation result for the
equimolar TiAl melt, comparing with experimental data obtained from
a zero scatterer.
We compare the simulation result to experimental data obtained by us
in Fig.~\ref{fig:SCC}.
In general, the simulation predicts a somewhat weaker ordering, expressed
through less pronounced oscillations in $S_{cc}(q)$.
The
temperature dependence in experiment is relatively weak, so that we
conclude that the experiment shows slightly stronger chemical ordering
than the simulation.
We also note that the soft-sphere model is in qualitative disagreement
with the observed $S_{cc}(q)$; this highlights, that the conclusions that
can be drawn from just the peak positions in $g(r)$ are limited.

\begin{figure}
    \centering
    \includegraphics[width=\linewidth]{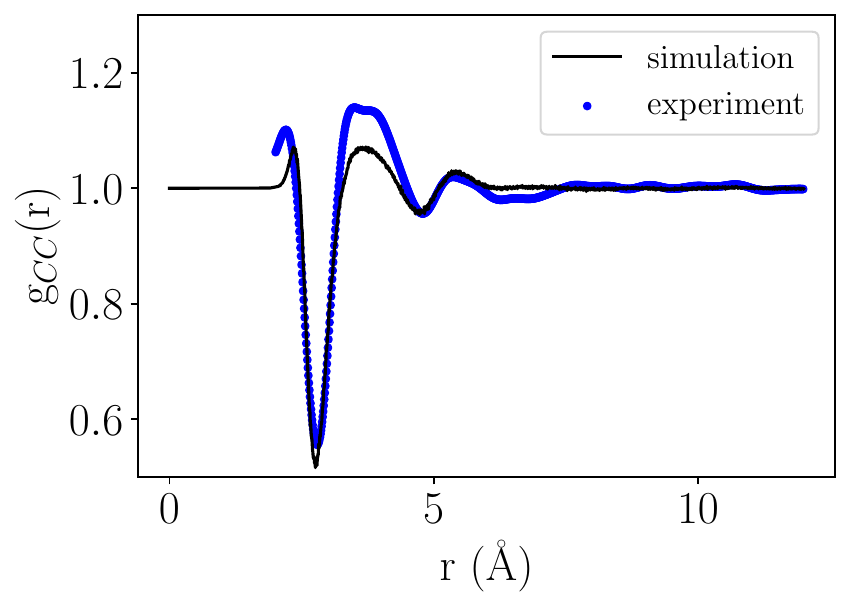}
    \caption{$g_{CC}(r)$ for equimolar Al-Ti at $T=1785K$ for simulations and experiment. }
    \label{fig:gcc}
\end{figure}

The real-space equivalent of $S_{cc}(q)$ shows more clearly the structural
charge-ordering in the mixture. In analogy to Eq.~\eqref{eq:hq}, we set
$S_{cc}(q)=x_\text{Al}x_\text{Ti}+nx_\text{Al}x_\text{Ti}h_{cc}(q)$,
which implies
\begin{equation}
g_{cc}(r)=1+x_\text{Al}x_\text{Ti}\left(g_\text{AlAl}(r)
  +g_\text{TiTi}(r)-2g_\text{AlTi}(r)\right)\,.
\end{equation}
Figure~\ref{fig:gcc} shows this quantity from the simulation (directly
calculated from the \glspl{rdf})
and from experiment (calculated
as the inverse Fourier transform of $(S_{cc}(q)/x_\text{Al}x_\text{Ti}-1)/n$).
One observes in $g_{cc}(r)$ a pronounced minimum that indicates preferential
ordering where atoms of uneqal species are close.
As expected from the data on $S_{cc}(q)$, the simulation underestimates
the extent of the ordering compared to the experiment, in particular in the
excess part of $g_{cc}(r)$ around $q\approx\qty{4}{\angstrom}$, and the
shape of the corresponding peak.

\subsection{Viscosity}
\label{sec:viscosity}
The \gls{md} simulation allows to connect the composition dependence of the
transport coefficients and thermophysical properties to the changes in
microscopic structure.
For the validation of the \gls{mlip}, a comparison to dynamical data
is crucial, since even reasonably good agreement in the static properties
can lead to inaccurate predictions of the dynamics.

\begin{figure}
    \centering
    \includegraphics[width=\linewidth]{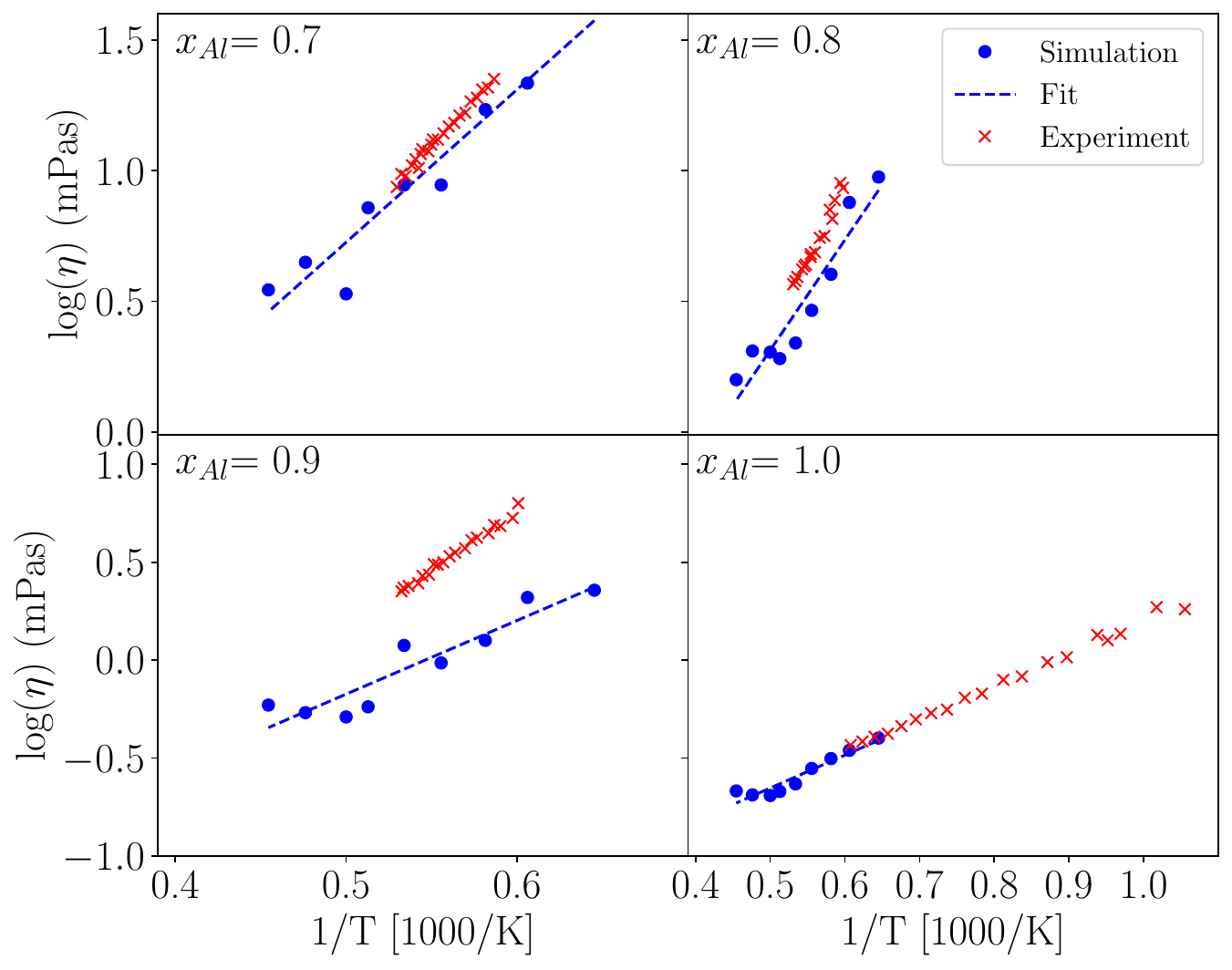}
    \caption{Temperature dependence of the viscosity of Al-Ti melts for different concentrations as labeled. Circles correspond to simulation results, red crosses to experimental data points taken from Ref.~\cite{TakedaSatoITJ2025}.
    Dashed lines indicate Arrhenius-like temperature dependencies from
    fits to the simulation data.
    \label{fig:visc-temp}
    }
\end{figure}

Figure~\ref{fig:visc-temp} shows the temperature-dependent viscosity for
compositions with $x_\text{Al}\ge0.7$, where experimental data is available
\cite{TakedaSatoITJ2025}.
(Experimental data is also available for $x_\text{Al}=0.6$ and shows very
similar agreement as the $0.7$ case that we discuss here.)

In the temperature range covered by the simulation and experiments,
the viscosity closely follows an Arrhenius
law, as corresponding fits (dashed lines in the figure) show.
While the viscosity values agree reasonably well between simulation and
experiment, systematic deviations become noticable: in particular, the
predicted activation energies, and consequently also the viscosity itself,
are predicted too low by the \gls{mlip}-based simulation. This becomes
most pronounced for $x_\text{Al}=0.9$,
and would thus suggest an increasingly
larger error with increasing Al concentration. However, the simulations
for pure Al again show much better agreement with the experiment.

The reason for this discrepancy is unclear. Recall that the density of the
system agrees very well with experiment (Fig.~\ref{fig:vd_00}), and this
agreement becomes systematically better as $x_\text{Al}\to1$. This suggests
to rule out a systematic density effect, where a lower density would also
predict a lower viscosity. It also highlights the importance of validating
the \gls{mlip} also against dynamical data.

\begin{figure}
    \centering
    \includegraphics[width=\linewidth]{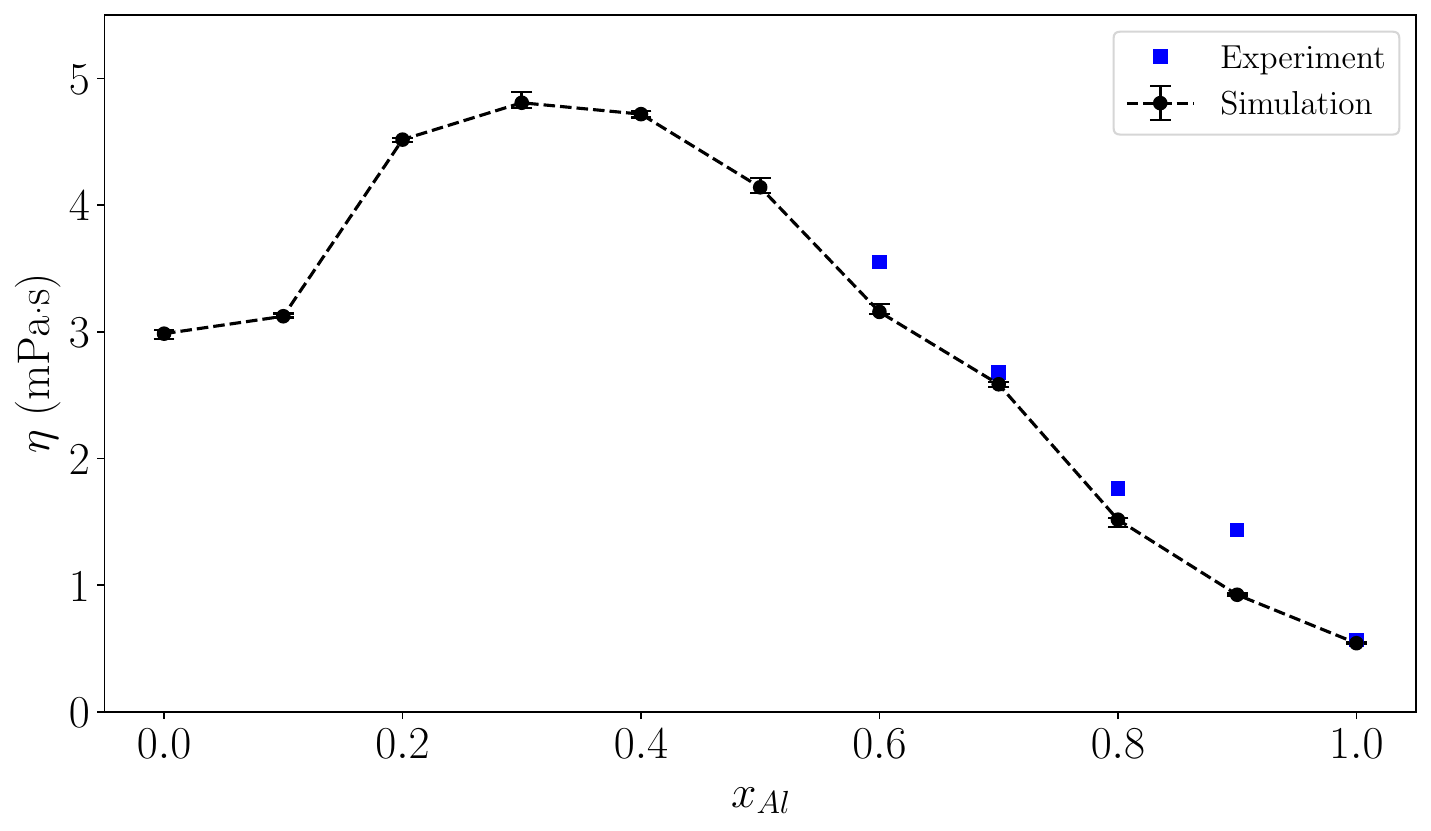}
    \caption{Viscosity of the Al-Ti melts as a function of Al concentration,
    for fixed temperature $T=\SI{1873}{\kelvin}$. Black symbols connected by
    dashed lines are the simulation results, blue squares represent
    experimental data from Ref.~\protect\cite{TakedaSatoITJ2025}.
    \label{fig:vs_01}
    }
\end{figure}

To investigate more clearly the concentration dependence of the viscosity,
we show results at a fixed temperature of $T=\SI{1873}{\kelvin}$ in Fig.~\ref{fig:vs_01}.
This representation emphasizes, that the \gls{mlip}-based simulation
predicts the correct trend of the viscosity with mixing: the viscosity
of pure Al increases with the increasing admixture of Ti in the range where
experimental data are available ($x_\text{Ti}\le0.4$).

The simulation predicts the isothermal variation of the viscosity to
undergo a maximum at an intermediate composition; for the temperature shown,
this occurs around $x_\text{Al}\approx0.3$.
Such a mixing-induced slowing down of the fluid is commonly found in
binary alloy systems, in particular in those with strong negative
excess Gibbs energy,
such as 
Al-Ni \cite{novakovic2024viscosity},
Al-Cu \cite{schick2012viscosity}, Al-Au \cite{peng2015structural}, or AlCuAg \cite{BrilloBook2016}. 

However, note that the maximum in the viscosity, around $x_\text{Al}\approx0.3$,
does not correlate with the
minimum in the excess volume (Fig.~\ref{fig:vd_11}), that simulation and
experiment consistently observe around $x_\text{Al}\approx0.6$.

\subsection{Diffusion coefficient}
\label{sec:results_diffusioncoefficient}

\begin{figure}
    \centering
     \includegraphics[width=\linewidth]{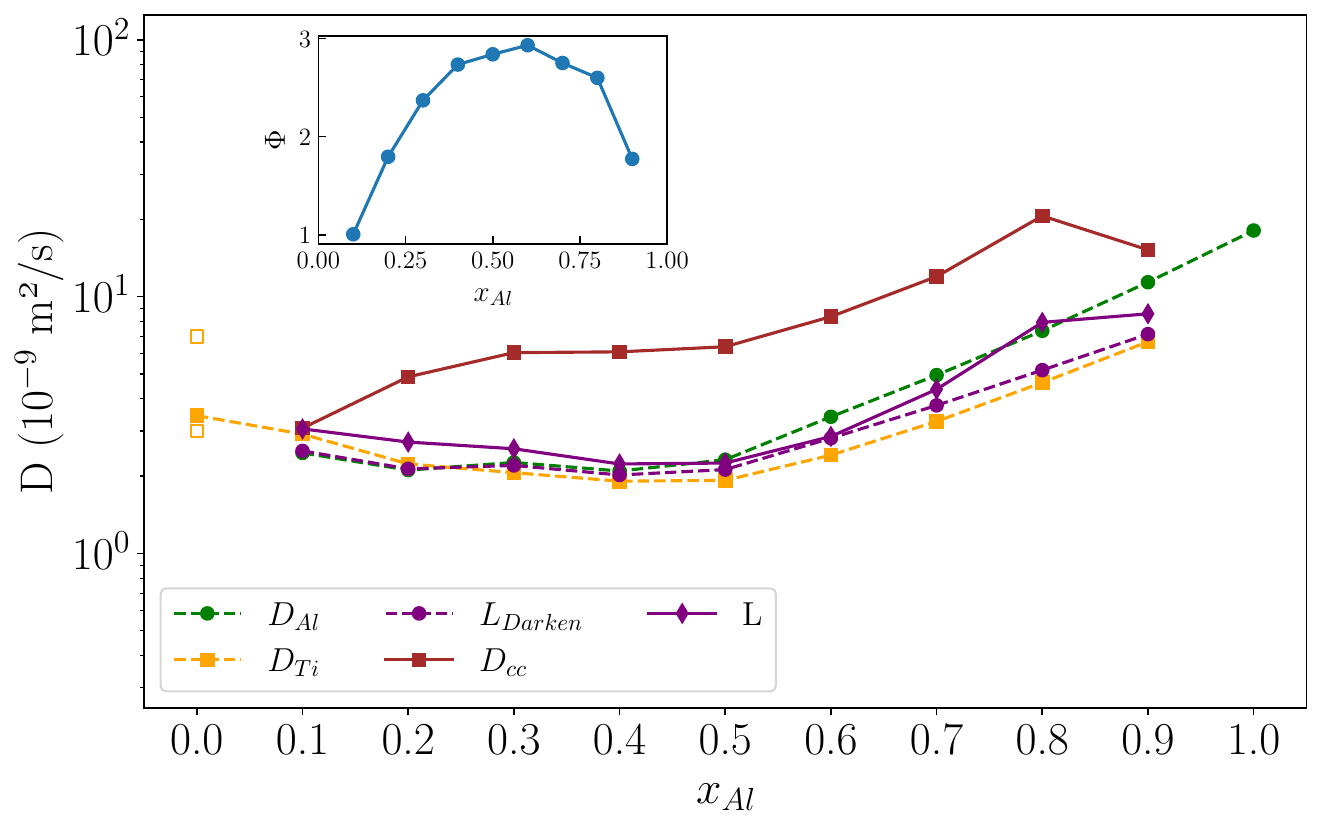}
    \caption{Concentration-dependent diffusion coefficients of the Al-Ti
    melt at $T=\qty{1720}{\kelvin}$. $D_{Al}$, $D_{Ti}$, and $D_{cc}$ are calculated with the corresponding MSDs. 
    $L$ is defined as $D_{cc}/ \Phi$ and the Darken approximation uses
    $L_\text{Darken} \approx x_{Ti}D_{Al} + x_{Al}D_{Ti} $. Unfilled symbols correspond to data from \cite{horbach2009ti}, using two different empirical simulation models (see text for details). Inset: $S_{cc}(q\to0)$, which is used to determine the thermodynamic factor. 
    \label{fig:md_00}
    }
\end{figure}

Figure~\ref{fig:md_00} shows the self-diffusion coefficients $D_\text{Al}$
and $D_\text{Ti}$ as functions of
concentration. As expected from the discussion of the viscosity
-- the melt becomes less fluid at fixed temperature upon mixing --
and the fact that an increased viscosity should result in a reduced
mobility of the constituent atoms,
the self-diffusion coefficients show broad minima around
$x_\text{Al}\approx0.3$. 

Experimental data for self-diffusion in liquid Ti exists only at
higher temperatures \cite{horbach2009ti}, but this data has been used to
gauge an empirical \gls{eam}. The original \gls{eam} potential published
by \Citeauthor{zope2003interatomic} \cite{zope2003interatomic} somewhat overestimates
the experimental values, and has been adjusted by \Citeauthor{horbach2009ti}
\cite{horbach2009ti} with a simple energy-rescaling factor to match the
experimental data. The two \gls{eam} results are shown in
Fig.~\ref{fig:md_00} at the composition $x_\text{Al}=0$ (upper and
lower open symbol, respectively).
Remarkably, the current \gls{mlip} very closely matches the value
of the adjusted \gls{eam}, which in turn accurately describes the
high-temperature diffusion data; in the case of the \gls{mlip} no further
adjustment was needed.

Figure~\ref{fig:md_00} also shows the interdiffusion coefficient $D_{cc}$.
As discussed above, $D_{cc}$ consists of a thermodynamic prefactor $\Phi$,
and a kinetic contribution $L$. These contributions are shown separately
in the figure. We observe a maximum in $\Phi$ at the fixed temperature
considered that occurs around $x_\text{Al}\approx0.6$. This corresponds
well to to the extremum seen in the excess volume, cf.\
Fig.~\ref{fig:vd_11}, but is quite distinct from the extremum seen
in the self-diffusion coefficients. The kinetic contribution to
interdiffusion, expressed through the Onsager coefficient $L$,
on the other hand shows a broad minimum at intermediate concentrations.
Thus the resulting interdiffusion coefficient is a combination of two
competing effects -- slowed-down kinetics but enhanced free-energy-induced
driving force. This is a generic effect that has been seen in different
alloy melts, for example in Al-Ni and Zr-Ni \cite{kuhn2014}.

The Darken approximation, Eq.~\eqref{eq_interdiff_darken}, presumes
that the kinetic contribution to interdiffusion is effectively governed
by the self-diffusion processes. We observe deviations from this
approximation in particular in the Al-rich melt:
$L_\text{Darken}$ is about a factor of $1.5$ smaller than the true
Onsager coefficient around $x_\text{Al}=0.8$, leading to a corresponding
underestimation of the interdiffusion coefficient by the Darken approximation.
Note that for $x_\text{Al}\to1$, there holds $D_{cc}\to D_\text{Ti}$,
and this value is extrapolated to be significantly smaller than the
Al self-diffusion coefficient. This for large Al concentrations,
interdiffusion becomes slower than Al-self diffusion.
On the Ti-rich side, $L$ is larger than both self-diffusion coefficients,
indicating that concentration gradients are relaxed collectively before
individual particles diffuse. This is different on the Al-rich side,
and we note that the crossover coincides with the point of strongest
negative excess volume around $x_\text{Al}\approx0.6$.

\begin{figure}
    \centering
    \includegraphics[width=\linewidth]{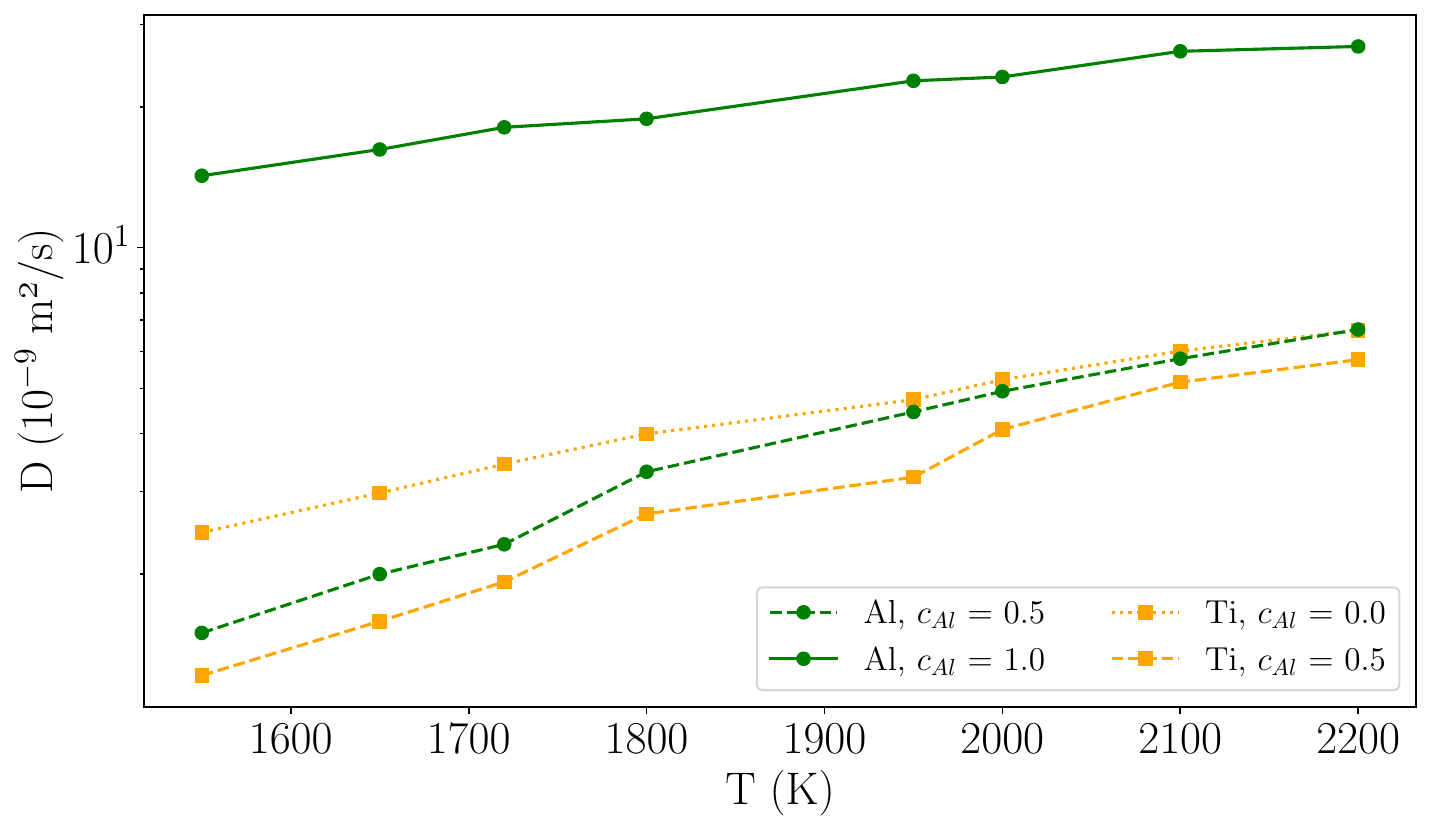}
   \caption{Diffusion coefficients as a function of temperature. Green lines represent $D_{\mathrm{Al}}$, orange lines represent $D_{\mathrm{Ti}}$. Solid lines correspond to $x_{\mathrm{Al}}=1$, dashed lines to $x_{\mathrm{Al}}=0.5$, and dotted lines to $x_{\mathrm{Al}}=0$.}
    \label{fig:diffusion_temperature}
\end{figure}

The temperature dependence of the self-diffusion coefficients is shown
in Fig.~\ref{fig:diffusion_temperature}. Besides the expected temperature
variation, it shows that self-diffusion in liquid Al
is almost one order of magnitude faster than self-diffusion in liquid Ti
at any given temperature.
The equimolar binary mixture displays diffusion coefficients that are dominated
by the slower Ti dynamics, as can also be seen from Fig.~\ref{fig:md_00}:
for small $x_\text{Al}$, the diffusion coefficients show little change
with composition and are close to the self-diffusion coefficient in pure Ti.
Only for $x_\text{Al}>0.5$, both self-diffusion coefficients strongly
increase with increasing Al concentration, approaching the faster
self-diffusion value of the pure Al melt.
The equimolar mixture also shows a more pronounced decrease in diffusion
as temperature is lowered, compared to the pure liquids,
in Fig.~\ref{fig:diffusion_temperature}.

\begin{figure}
    \centering
    \includegraphics[width=\linewidth]{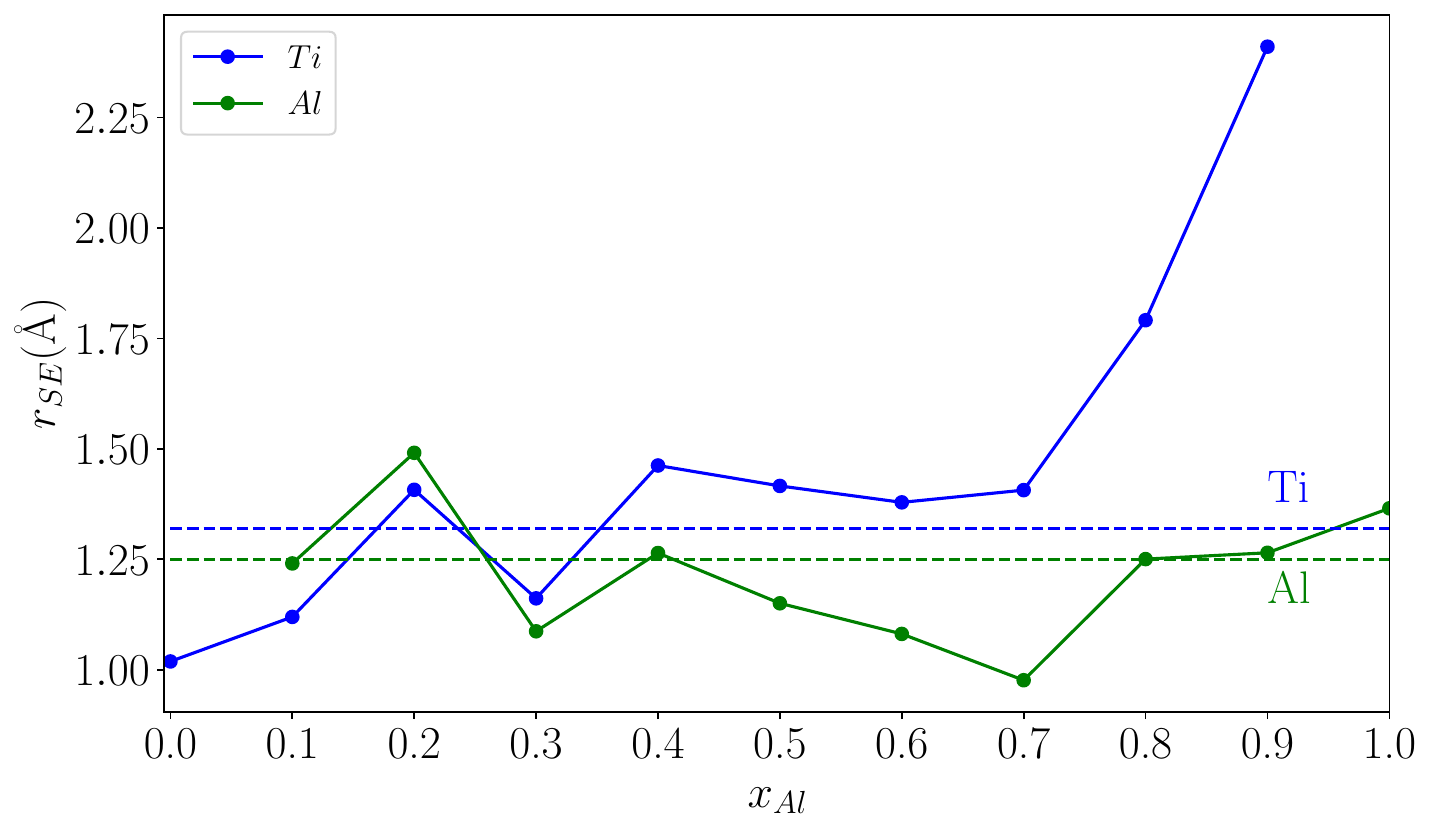}
    \caption{Concentration-dependent hydrodynamic radius $r_\text{SE}$ from the Stokes--Einstein (SE) relation of the Al-Ti
    melt at $T=\qty{1873}{\kelvin}$. Horizontal lines correspond to the Pauling atomic radii taken from Ref.~\cite{slater1964atomic},
    \textit{i.e.}, $r_\text{Al}=\qty{1.25}{\angstrom}$ and
    $r_\text{Ti}=\qty{1.32}{\angstrom}$. 
    \label{fig:rSE}
    }
\end{figure}

The notion that the mobility of individual atoms is a proxy for the
fluidity of the mixture, and hence the inverse viscosity, is reflected
in the empirical \gls{se} relation. Borrowed from the famous result for
the diffusion of a large sphere in a continuum solvent, one
expects a relation of the form
\begin{equation}\label{eq:se}
  D_\alpha\eta=\frac{k_BT}{\alpha r_\alpha}\,
\end{equation}
where $\alpha=6\pi$ would hold for a large sphere of radius $r_\alpha$
embedded in a solvent of viscosity $\eta$ with stick boundary
conditions on the sphere's surface. For molecular fluids, this relation
does not necessarily hold, although a strong coupling of $D_\alpha^{-1}$
and $\eta$ is often observed \cite{BrilloPRL2011}. The prefactor $\alpha$ is debatable,
and from kinetic theory, $\alpha=4\pi$ is suggested, although $\alpha=6\pi$
is more commonly taken in the literature.

Equation~\eqref{eq:se} allows to define an effective \gls{se} radius
of the particles in the melt. Taking $\alpha=6\pi$,
\begin{equation}
    r_{\text{SE},\alpha} = \frac{k_\mathrm{B} T}{6 \pi  D_\alpha \eta}
    \label{eq:SErelation}
\end{equation}
We show the values obtained from the simulation at $T=\qty{1873}{\kelvin}$
in Fig.~\ref{fig:rSE}. If the \gls{se} relation were accurate, the
observed values should be constant in a mixture with small chemical-mixing
effects; deviations are known to set in if the temperature is decreased
towards the melting point \cite{BrilloAPL2008}.
In Fig.~\ref{fig:rSE} one notes a strong composition dependence in
particular for $r_{\text{SE},\text{Ti}}$ in the Al-rich melt. This reflects
the observation that Ti-self-diffusion in the Al-rich melt is strongly
enhanced compared to the Ti-rich melts (see Fig.~\ref{fig:md_00}):
Assuming that in the Al-rich melt, viscosity and Al-diffusion are strongly
coupled, the lower viscosity results in an enhanced \gls{se} radius.

On the Ti-rich side, the \gls{se} radii fluctuate around values that are
compatible with the literature values for the covalent (Pauling)
atomic radii \cite{slater1964atomic},
$r_{c,\text{Al}}\approx\qty{1.25}{\angstrom}$ and
$r_{c,\text{Ti}}\approx\qty{1.32}{\angstrom}$.
Note however that these are not the values that best describe the
maxima in the $g_{\alpha\beta}(r)$.
For these structural properties,
\Citeauthor{miracle2010} \cite{miracle2010} have proposed somewhat larger radii,
$r_{M,\text{Al}}\approx\qty{1.41}{\angstrom}$ and
$r_{M,\text{Ti}}\approx\qty{1.42}{\angstrom}$;
even larger are the Goldschmidt (or metallic) radii,
$r_{m,\text{Al}}\approx\qty{1.432}{\angstrom}$ and
$r_{m,\text{Ti}}\approx\qty{1.448}{\angstrom}$.
Note that the values that enabled the soft-sphere model to describe
the peak positions in $g_{\alpha\beta}(r)$ reasonably well, are a mix
of these values: the value $r_{s,\text{Al}}\approx1.35$ is somewhat
larger than the covalent radius, while $r_{s,\text{Ti}}\approx1.448$ is
close to the Goldschmidt radius.
This shows that the ambiguity of defining atomic radii is not just confined
to any specific element, but can also depend strongly on the chemical
environment.

\section{Conclusion}
\label{sec:conclusion}
\glsresetall

We have investigated thermophysical and dynamical properties of
Al-Ti melts through \gls{md} simulations using a \gls{mlip},
the \gls{nep} potential NEP89 \cite{liang2025nep89}.
The simulations were validated against experimental data where available,
in particular for the composition-dependent density, the static
structure factor in the case of liquid Ti, the $S_{cc}(q)$ in the case of
Al-Ti, and the
temperature-dependent viscosities at experimentally available
Al-rich compositions.

The \gls{mlip} shows in general good predictive capability for the
Al-Ti melts. Note that this potential was never specifically trained
for this system, and presumably also with a focus on solid-state properties
as implicit in the large ab-initio datasets that are the basis of
the \gls{nep} potential. This shows that this modern \gls{mlip}
has a high degree of transferability.

The Al-Ti melts show, similar to other Al-based alloy systems, a pronounced
negative excess volume, which is well reproduced in the simulation.
Despite this strong indication of non-ideal mixing, we found from an
analysis of the microscopic static structure, that mixing effects in
Al-Ti are predominantly substitutional, according to the microscopic
structure functions.
While a simple hard-sphere model is not sufficient
to explain the concentration dependence of the local packing, the latter
can be very well described by a soft-sphere model; this explains a
mixing-induced
``local compression'' effect that arises from the soft interactions.
However, chemical short-range order is present, as evidenced by a
non-trivial $S_{cc}(q)$ in simulation that matches well with experiment,
but is not matched by the soft-sphere model.

The most pronounced deviations in the static properties were seen in the
density on the Ti-rich side; their reason is unclear and potentially
attributed to the simulation model, but potenbtially also to
systematic errors in experiment (such as contamination with a
foreign species). In principle, the effect of impurities could be
studied in further \gls{mlip}-based simulations; we leave this for
future work.

Regarding the dynamical properties, we found general good agreement in the
prediction of the viscosities, but deviations at a composition of
$x_\text{Al}=90\%$ whose origin is unclear. This coincides with an
under-estimation in the simulation of the activation energy for the
Arrhenius-like dependence of viscosity on temperature. This is not
obviously related to a deficiency of the potential in describing the
microscopic structure of the melt. It thus highlights, that accurate
experimental data for the transport coefficients in the melt are an
important (but currently often under-valued) further ingredient to
validate machine-learned descriptions of alloys.
This becomes in particular important for strongly correlated liquids
where a strong increase with viscosity as a function of decreasing
temperature is seen, and should affect studies employing \gls{mlip}
for the simulations of glass-forming materials \cite{pabst2025glassy}.

We note that the concentration-dependence of the viscosity under
isothermal conditions shows excellent agreement between simulation
and experiment. This gives confidence in the \gls{mlip}-based
predictions of the self- and inter-diffusion coefficients for which
there are no experimental data yet. We observe isothermal mixing
effects that are quite common for binary alloys: upon mixing, the
viscosity displays a maximum at some intermediate composition,
while the diffusion coefficients show a corresponding minimum.
The interdiffusion process on the contrary is a competition between a
strong thermodynamic driving force, expressed through a thermodynamic
factor larger than unity, and a kinetics-induced minimum in the
Onsager coefficient. Especially on the Al-rich side, the commonly used
Darken approximation for the interdiffusion coefficient fails.
Also the Stokes-Einstein relation breaks down in the composition range,
and we attribute this to the fact that the Al-rich melt shows collective
dynamics that is much more fluid than that of the Ti-rich melt, hence
giving rise to a decrease in viscosity that is not expressed in the
Ti-tracer dynamics.

\begin{acknowledgments}
Y.K. acknowledges financial support from the Dentsu Ikueikai Foundation.
The authors gratefully acknowledge the scientific support and HPC resources
provided by the German Aerospace Center (DLR). The HPC system CARO is
partially funded by ``Ministry of Science and Culture of Lower Saxony'' and
``Federal Ministry for Economic Affairs and Climate Action''.
Part of this work was funded by the Deutsche Forschungsgemeinschaft (DFG,
German Research Foundation) -- project number 505695238, part of the
ANR-DFG project SOLIMAT.
\end{acknowledgments}

\section*{Data availability}
The data that support the findings of this article are openly available \cite{zenodo1}. 

\bibliography{sample}

\end{document}